 \documentclass[aps,prd,preprintnumbers,superscriptaddress,nofootinbib,showpacs,twocolumn]{revtex4}%
\usepackage[dvipdfmx]{graphicx}
\usepackage{bm,latexsym,amsmath,amssymb,amsfonts,mathrsfs}
\usepackage{color}
\input{colordvi.tex}
\allowdisplaybreaks[1]
\usepackage[dvipdfmx]{hyperref}
\usepackage{url}
\usepackage{ulem}
\usepackage{hyperref}
\hypersetup{
 colorlinks=true,
  citecolor=cyan,
 }
\newcommand*{\D}{{\rm d}}
\newcommand*{\mpl}{M_{\rm Pl}}

\begin{document}

\title{Imprints of primordial gravitational waves with non-Bunch-Davies initial states on CMB bispectra}
\author{Shingo~Akama}
\email[Email: ]{shingo.akama"at"uj.edu.pl}
\affiliation{Faculty of Physics, Astronomy and Applied Computer Science, Jagiellonian University, 30-348 Krakow, Poland
}
\author{Hiroaki~W.~H.~Tahara}
\email[Email: ]{tahara"at"rikkyo.ac.jp}
\affiliation{Department of Physics, Rikkyo University, Toshima, Tokyo 171-8501, Japan
}

\begin{abstract}
It has been shown that both scalar and tensor modes with non-Bunch-Davies initial states can enhance the amplitudes of the primordial bispectra compared to those with the Bunch-Davies state, especially for wavenumber modes in a flattened triangle configuration. However, in the case of the non-Bunch-Davies scalar modes, it has also been found that those enhancements in Fourier space are somewhat reduced in bispectra of cosmic microwave background (CMB) fluctuations. In this paper, we show that the enhancement resulting from the tensor modes is partially reduced to a degree differing from that of the scalar modes, which makes the non-Bunch-Davies effects unobservable in gravitational theories with the same quadratic and cubic operators of the tensor perturbations as general relativity. Furthermore, we present examples of gravitational theories yielding enhancements that would potentially be detected through CMB experiments.

\end{abstract}

\preprint{RUP-23-12}

\maketitle
\section{Introduction}

Inflation~\cite{Guth:1980zm,Sato:1980yn,Linde:1983gd} is regarded as a successful paradigm for the early universe, providing an elegant resolution for various problems in the standard Big-Bang cosmology and the mechanism behind the generation of the rich structures of our universe. In the future, it is expected that cosmological observations will clarify which model of inflation aligns best with the actual early universe.

Correlations in the cosmic microwave background (CMB) are anticipated to serve as distinctive fingerprints for differentiating inflation models, and this has been a focus of extensive research in the past~\cite{WMAP:2010qai,Planck:2018vyg,Planck:2018jri}. While temperature correlations have been observed with remarkable precision, confirming nearly perfect Gaussian power spectra and very small non-Gaussianity, the statistics of polarization remains less certain. Significant improvements in the accuracy of future observations are anticipated.

In this context, the study of non-Gaussianity stands at the frontier of cosmology. The detection of higher-order spectra, such as the bispectrum of polarization, can be regarded as the final objective in CMB observations. Specifically, given that B-mode polarization is not generated from curvature fluctuations but solely from gravitational waves~\cite{Seljak:1996gy}, the statistical nature of B-mode polarization constitutes a vital area of research. This holds the potential to significantly contribute to our understanding of the physics of the extremely early universe.

In this paper, we study the effects of the tensor modes with non-Bunch-Davies initial conditions on the bispectra of the CMB fluctuations. The primordial power spectra and bispectra associated with the non-Bunch-Davies states have been comprehensively studied~\cite{Chen:2006nt,Holman:2007na,Xue:2008mk,Meerburg:2009ys,Meerburg:2009fi,Chen:2010xka,Meerburg:2010ks,Chen:2010bka,Meerburg:2010rp,Agullo:2010ws,Ashoorioon:2010xg,Ganc:2011dy,LopezNacir:2011kk,Ganc:2012ae,Agarwal:2012mq,Gong:2013yvl,Flauger:2013hra,Aravind:2013lra,Ashoorioon:2013eia,Ashoorioon:2014nta,Brahma:2013rua,Bahrami:2013isa,Kundu:2013gha,Emami:2014tpa,Zeynizadeh:2015eia,Meerburg:2015yka,Ashoorioon:2016lrg,Shukla:2016bnu,Brahma:2018hrd,Ashoorioon:2018sqb,Brahma:2019unn,Akama:2020jko,Naskar:2020vkd,Ragavendra:2020vud,Fumagalli:2021mpc,Kanno:2022mkx,Ghosh:2022cny,Gong:2023kpe}. In particular, in Refs.~\cite{Chen:2006nt,Holman:2007na,Xue:2008mk,Meerburg:2009ys,Meerburg:2009fi,Chen:2010bka,Ashoorioon:2010xg,Bahrami:2013isa,Akama:2020jko,Naskar:2020vkd,Kanno:2022mkx,Ghosh:2022cny,Gong:2023kpe}, it has been demonstrated that the primordial bispectra for nearly-flattened triangles can be enhanced in the presence of the modes that deviate from the Bunch-Davies state. However, it has also been found in Ref.~\cite{Holman:2007na} that a part of the enhancements in the primordial scalar bispectra are reduced due to the necessary angular average when deriving the CMB bispectra. This reduction has not been discussed in the context of the tensor non-Gaussianity yet. In addition, differently from the scalar non-Gaussianities, the tensor ones for the exactly flattened triangle vanish, which has been shown in Ref.~\cite{Akama:2020jko}. (See also Ref.~\cite{Gong:2023kpe} for some debate about subtlety.) Therefore, it is important to investigate how the enhancements around the flattened triangles are reduced in the observable quantities originating from the tensor non-Gaussianities.

The primordial non-Gaussianities that peak around flattened triangles are considered to be generated on subhorizon scales. Thus, one might naively think that such non-Gaussianities would be amplified more if there were cubic operators involving higher derivatives. The cubic operator is unique in a certain class of gravitational theories (e.g., general relativity with a canonical scalar field), whereas additional cubic operators with higher derivatives are in some extended theories (e.g., the Horndeski theory, which gives the most general second-order field equations for a scalar field and a metric~\cite{Horndeski:1974wa,Deffayet:2011gz,Kobayashi:2011nu}, and its generalized theory called Gleyzes-Langlois-Piazza-Vernizzi (GLPV) theory~\cite{Zumalacarregui:2013pma,Gleyzes:2014dya,Gleyzes:2014qga}). So far, tensor non-Gaussianities in such extended theories have been analyzed in a unified way within the GLPV theory by Ref.~\cite{Akama:2020jko}. In this paper, in the same way as in the literature, we first investigate the impact of tensor non-Gaussianities on CMB bispectra in the GLPV theory. Subsequently, we also see whether a further extension of a gravitational theory with higher-derivative cubic operators would result in further enhanced CMB bispectra or not.

This paper is outlined as follows. In the following section, we introduce our setup to describe scalar and tensor perturbations with non-Bunch-Davies initial states. In Sec.~\ref{sec: reduction}, we first review the reduction of the enhancement of the scalar non-Gaussianities in the process of angular average. Then we extend the method used for the scalar non-Gaussianities to the tensor ones and clarify how the enhancements are reduced. In Sec.~\ref{sec: CMB-bispectra-GLPV-beyond}, we evaluate the actual enhancements in CMB bispectra in GLPV and beyond-GLPV theories, taking into account the aforementioned reduction. We summarize this paper in Sec.~\ref{sec: summary}.

\section{Setup} 
We begin with a spatially-flat Friedmann-Lema\^{i}tre-Robertson-Walker (FLRW) spacetime and employ the following Arnowitt-Deser-Misner (ADM) metric:
\begin{align}
\D s^2=-N^2\D t^2+\gamma_{ij}(\D x^i+N^i\D t)(\D x^j+N^j\D t),
\end{align}
where $N=1, N_i=0$, and $\gamma_{ij}=a^2\delta_{ij}$ with $a$ denoting the scale factor at the background level. Throughout this paper, we consider de Sitter inflation models where the scale factor is $a\simeq-1/(H\eta)$ with the Hubble parameter $H:=({\rm d}a/{\rm d}t)/a\simeq{\rm const.}$ and conformal time $\eta$. The perturbed variables are defined in the unitary gauge, $\phi(t,\vec x)=\phi(t)$, as
\begin{align}
N&=1+\delta n, N_i=\partial_i\chi,\\
\gamma_{ij}&=a^2e^{2\zeta}\biggl(\delta_{ij}+h_{ij}+\frac{1}{2}h_{ik}h^k_j+\cdots\biggr),
\end{align}
where one obtains the auxiliary fields $\delta n$ and $\chi$ by solving the constraint equations, and we denoted the curvature perturbation by $\zeta$ and the gravitational waves by $h_{ij}$.

Throughout this paper, we consider the non-Bunch-Davies initial states under which the quantized perturbations are expanded as
\begin{align}
\zeta(t,{\bf k})&=\zeta_k b_{\bf k}+\zeta^*_k b^\dagger_{-{\bf k}},\\
h_{ij}(t,{\bf k})&=\sum_s\left[\psi^{(s)}_k e^{(s)}_{ij}({\bf k})b^{(s)}_{{\bf k}}+\psi^{(s)*}_k e^{(s)*}_{ij}(-{\bf k})b^{(s)\dagger}_{-{\bf k}}\right],
\end{align}
where the transverse and traceless polarization tensor $e^{(s)}_{ij}({\bf k})$ satisfies $e^{(s)}_{ij}({\bf k})e^{(s')*}_{ij}({\bf k})=\delta_{ss'}$, the subscript $s$ denotes the two hecility modes of the gravitational waves, and $b^\dagger_{\bf k}$ ($b_{\bf k}^{(s)\dagger}$) and $b_{\bf k}$ ($b_{\bf k}^{(s)}$) stand for the creation and annihilation operators of the scalar modes (tensor modes), respectively. Also, those operators satisfy the canonical commutation relations:
\begin{align}
[b_{\bf k},b^\dagger_{\bf k'}]&=(2\pi)^3\delta({\bf k}-{\bf k}'),\\
[b^{(s)}_{\bf k},b_{\bf k'}^{(s')\dagger}]&=(2\pi)^3\delta_{ss'}\delta({\bf k}-{\bf k}'),\\
{\rm others}&=0.
\end{align}
The mode functions with the non-Bunch-Davies states are obtained from those with the Bunch-Davies one via Bogoliubov transformations. In particular, when $u_k$ and $v_k$ stand for the positive frequency mode function of the curvature perturbations and tensor perturbations, respectively, we have
\begin{align}
\zeta_k&=\alpha_k u_k+\beta_k u^*_k,\\
\psi^{(s)}_k&=\alpha^{(s)}_kv_k+\beta^{(s)}_kv^*_k,
\end{align}
where the Bogoliubov coefficients satisfy the following normalization conditions
\begin{align}
|\alpha_k|^2-|\beta_k|^2&=1,\\
|\alpha^{(s)}_k|^2-|\beta^{(s)}_k|^2&=1. \label{eq: normali-bogo}
\end{align}
The explicit forms of the mode functions depend on a concrete model, which will be defined later.

\section{Reduction of Non-Bunch-Davies effects}\label{sec: reduction}
If the initial perturbations were in the Bunch-Davies state, the solution of the mode function is represented by the positive frequency mode. In contrast, if they began in the non-Bunch-Davies states, the solution includes both positive and negative frequency modes. The interplay between these modes has been shown to yield enhancements of the primordial bispectra, the quantity calculated in Fourier space~\cite{Chen:2006nt,Holman:2007na,Xue:2008mk,Meerburg:2009ys,Meerburg:2009fi,Chen:2010bka,Ashoorioon:2010xg,Bahrami:2013isa,Akama:2020jko,Naskar:2020vkd,Kanno:2022mkx,Ghosh:2022cny,Gong:2023kpe}. However, it has been shown that the enhancements of the scalar auto-bispectrum are somewhat reduced in the CMB (temperature) bispectrum, the quantity obtained after projecting the primordial scalar bispectrum onto the two-dimensional celestial surface~\cite{Holman:2007na}. In the following subsection, we first review a method used in Ref.~\cite{Holman:2007na} to clarify the reduction for the scalar auto-bispectrum. Then we extend that method to the tensor auto-bispectrum and quantify the extent of reduction in enhancements.

\subsection{Scalar Bispectrum}
Let us review the analysis in Ref.~\cite{Holman:2007na}. The primordial bispectrum of the curvature perturbations $\mathcal{B}_\zeta$ is defined by
\begin{align}
\langle\zeta({\bf k}_1)\zeta({\bf k}_2)\zeta({\bf k}_3)\rangle=(2\pi)^3\delta({\bf k}_1+{\bf k}_2+{\bf k}_3)\mathcal{B}_\zeta.
\end{align}
The three-point correlation function can be obtained by following the in-in formalism. The authors of Ref.~\cite{Holman:2007na} considered the scalar-field model where the quadratic action of curvature perturbations is of the form,
\begin{align}
S^{(2)}_\zeta=\int\D t\D^3xa^3\biggl[\mathcal{G}_S\dot\zeta^2-\frac{\mathcal{F}_S}{a^2}(\partial_i\zeta)^2\biggr].
\end{align}
The mode function can be derived from the above as
\begin{align}
u_k=\frac{1}{\sqrt{2}a(\mathcal{G}_S\mathcal{F}_S)^{1/4}}\frac{\sqrt{\pi}}{2}\sqrt{-c_s\eta}H^{(1)}_{3/2}(-c_sk\eta),
\end{align}
where $H^{(1)}_{3/2}$ is the Hankel function of the first kind of order $3/2$ and $c_s^2$ is the square of the propagation speed of the curvature perturbations defined by $c_s^2:=\mathcal{F}_S/\mathcal{G}_S$. Here, it has been assumed that $\mathcal{G}_S, \mathcal{F}_S$ are constants. In this framework, the resultant primordial bispectrum includes the following term~\cite{Holman:2007na}:
\begin{align}
\mathcal{I}_s:=f_s(k_i)\int_{\eta_0}^0\D\eta(-\eta)^n e^{-ic_s\tilde k_j\eta}, \label{eq: int-scalar}
\end{align}
where $\tilde k_j:=-k_j+k_{j+1}+k_{j+2}$ with $j$ being defined modulo $3$ and $\eta_0$ is the conformal time when the perturbations are on subhorizon scales, $-c_sk_i\eta_0\gg1$. We assume that the theories considered in the present paper are valid up to the cutoff scale $\Lambda=k/a(\eta_0)\simeq(-k\eta_0)\cdot H$.
Note that, unlike in the case of the Bunch-Davies state where $\tilde k_j$ takes the value $k_j+k_{j+1}+k_{j+2}$, the coefficient of $k_j$ in $\tilde k_j$ has the opposite sign to the others due to mixing between the positive and negative frequency modes. Seen from the above equation, the condition $-c_s\tilde k_j\eta_0\ll1$ leads to a non-oscillating integrand, which produces a peak around $\tilde k_j=0$. We call $\tilde k_j=0$ (i.e., $k_j=k_{j+1}+k_{j+2}$) the exact flattened configuration.

Eq.~(\ref{eq: int-scalar}) is a key integral to evaluate the enhancement of the primordial scalar bispectrum and its reduction for the CMB bispectrum. The other terms in the primordial bispectrum are irrelevant to the arguments on the enhancement and the reduction, which we do not consider here. Eq.~(\ref{eq: int-scalar}) yields the term proportional to $(-k_i\eta_0)^{n+1}$ in the primordial bispectrum for the flattened triangle, whereas it has been shown in Ref.~\cite{Holman:2007na} that the CMB bispectrum receives $\mathcal{O}((-k_i\eta_0)^n)$-enhancement, i.e., one power of $(-k_i\eta_0)$ is reduced in the quantity observed by CMB experiments. In the following, we first review the loss of one power of $(-k_i\eta_0)$ in the CMB bispectrum originating from the curvature perturbations. 

The contribution from Eq.~(\ref{eq: int-scalar}) to the three-point correlation function of CMB fluctuations reads
\begin{widetext}
\begin{align}
\left\langle a_{l_1m_1}a_{l_2m_2}a_{l_3m_3}\right\rangle
&\ =(4\pi)^3(-i)^{l_1+l_2+l_3}\int\frac{{\rm d}^3{\bf k}_1}{(2\pi)^3}\frac{{\rm d}^3{\bf k}_{2}}{(2\pi)^3}\frac{{\rm d}^3{\bf k}_{3}}{(2\pi)^3}\mathcal{T}_{l_1}(k_1)\mathcal{T}_{l_{2}}(k_{2})\mathcal{T}_{l_{3}}(k_{3})\notag\\
&\quad \times Y^*_{l_1m_1}(\hat n_1)Y^*_{l_2m_{2}}(\hat n_{2})Y^*_{l_{3}m_{3}}(\hat n_{3})(2\pi)^3\delta({\bf k}_1+{\bf k}_2+{\bf k}_3)\mathcal{B}_\zeta\notag\\
&\ \supset(4\pi)^3(-i)^{l_1+l_2+l_3}\int\frac{{\rm d}^3{\bf k}_1}{(2\pi)^3}\frac{{\rm d}^3{\bf k}_2}{(2\pi)^3}\frac{{\rm d}^3{\bf k}_3}{(2\pi)^3}\mathcal{T}_{l_1}(k_1)\mathcal{T}_{l_2}(k_2)\mathcal{T}_{l_{3}}(k_{3})\notag\\
&\quad \times Y^*_{l_1m_1}(\hat n_1)Y^*_{l_{2}m_{2}}(\hat n_{2})Y^*_{l_{3}m_{3}}(\hat n_{3})(2\pi)^3\delta({\bf k}_1+{\bf k}_{2}+{\bf k}_{3})\mathcal{I}_s,
\end{align}
\end{widetext}
where $a_{lm}$ are the expansion coefficients of the CMB fluctuations in terms of the spherical harmonics, $\hat n_i$ are unit vectors defined by $\hat n_i:={\bf k}_i/|{\bf  k}_i|$ $(i=1,2,3)$, and $\mathcal{T}_l(k)$ denotes a transfer function of the temperature fluctuation or the E-mode polarization originating from the scalar perturbations.

In Ref.~\cite{Holman:2007na}, the manipulations for the argument on the reduction have been performed in an analytical way as follows. The time integral in $\mathcal{I}_s$ is sharply peaked at the flattened configuration, contrasting with other components such as $\mathcal{T}_l(k), Y^*_{lm}(\hat n)$, and $f_s(k_i)$. Consequently, the dominant contribution comes from the time integral in Eq.(\ref{eq: int-scalar}) for the flattened triangle. In light of this, after performing ${\bf k}_{j+2}$-integral for any $j$ via the delta function, the authors of Ref.~\cite{Holman:2007na} took the flattened limit of $\mathcal{T}_l(k), Y^*_{lm}(\hat n)$, and $f_s(k_i)$:
\begin{align}
&Y^*_{l_jm_j}(\hat n_j)Y^*_{l_{j+1}m_{j+1}}(\hat n_{j+1})Y^*_{l_{j+2}m_{j+2}}(\hat n_{j+2})\notag\\
&\to Y^*_{l_jm_j}(-\hat n_{j+1})Y^*_{l_{j+1}m_{j+1}}(\hat n_{j+1})Y^*_{l_{j+2}m_{j+2}}(\hat n_{j+1}),
\end{align}
and $k_{j+2}=|{\bf k}_j-{\bf k}_{j+1}|\to|k_j-k_{j+1}|$ in $\mathcal{T}_l(k)$ and $f_s(k_i)$.
Then, they performed the angular integral with respect to $\hat n_j$ (the angle between ${\bf k}_j$ and ${\bf k}_{j+1}$) as
\begin{widetext}
\begin{align}
\int\D^2\hat n_j I_s(\tilde k_j)=2\pi\int_{\eta_0}^0\D\eta(-\eta)^n\frac{1}{c_s^2k_jk_{j+1}\eta^2}[e^{2ic_sk_{j+1}\eta}(1-ic_s(k_j+k_{j+1})\eta)-(1-ic_s(k_j-k_{j+1})\eta)]\propto(-\eta_0)^n,
\end{align}
\end{widetext}
indicating that one power of $|k\eta_0|$ in the primordial bispectrum is diminished in the CMB bispectrum since $\mathcal{B}_\zeta\propto(-\eta_0)^{n+1}$. Indeed, similar reductions in the exponent have been reported in numerical calculations given in a previous study~\cite{Meerburg:2009ys}. In the following subsection, we adopt a similar analytical approach to evaluate the three-point function of the tensor perturbations.

\subsection{Tensor Bispectrum}

The three-point function of the tensor perturbations is defined by
\begin{align}
\langle\xi^{(s_1)}({\bf k}_1)\xi^{(s_2)}({\bf k}_2)\xi^{(s_3)}({\bf k}_3)\rangle=(2\pi)^3\delta({\bf k}_1+{\bf k}_2+{\bf k}_3)\mathcal{B}_h,
\end{align}
where $\xi^{(s)}({\bf k}):=h_{ij}(t,{\bf k})e^{(s)*}_{ij}({\bf k})$, and $\mathcal{B}_h$ is the primordial bispectrum of the tensor perturbations.
Throughout this paper, we focus on the gravitational theories that yield the standard form of the quadratic action:
\begin{align}
S^{(2)}_h&=\int\D t\D^3xa^3\biggl[\mathcal{G}_T\dot h_{ij}^2-\frac{\mathcal{F}_T}{a^2}(\partial_k h_{ij})^2\biggr]. \label{eq: quad-tensor}
\end{align}
General relativity is the simplest example giving this quadratic action with $\mathcal{G}_T=\mathcal{F}_T=\mpl^2$. The Horndeski theory which yields the most general second-order field equations~\cite{Horndeski:1974wa,Deffayet:2011gz,Kobayashi:2011nu} and the GLPV theory~\cite{Gleyzes:2014dya,Gleyzes:2014qga} also have Eq.~\eqref{eq: quad-tensor} as their quadratic action. (See also Ref.~\cite{Kobayashi:2019hrl} for a review for these frameworks.) Furthermore, we will also consider a beyond-GLPV class having the above quadratic action in this paper.

In this framework, the mode function is obtained as
\begin{align}
v_k=\frac{2}{a(\mathcal{G}_T\mathcal{F}_T)^{1/4}}\frac{\sqrt{\pi}}{2}\sqrt{-c_h\eta}H^{(1)}_{3/2}(-c_hk\eta),
\end{align}
where $c_h^2:=\mathcal{F}_T/\mathcal{G}_T$, and we assumed that $\mathcal{G}_T$ and $\mathcal{F}_T$ are constant similarly to $\mathcal{G}_S$ and $\mathcal{F}_S$. In the present setup, the primordial tensor auto-bispectrum is proportional to $\tilde k_j$ and vanishes for $\tilde k_j=0$.\footnotemark
(See Ref.~\cite{Akama:2020jko} and Appendix~\ref{sec: App}.) By taking this into account, we apply a similar argument as Eq.~(\ref{eq: int-scalar}) to the tensor mode and get
\begin{align}
\mathcal{I}_h:=f_h(k_i,s_i)\int_{\eta_0}^0\D\eta c_h\tilde k_j(-\eta)^n e^{-ic_h\tilde k_j\eta}, \label{eq: int-tensor}
\end{align}
where $i=1,2,3$ and $f_h\neq0$ for $\tilde k_j=0$. 
Since the explicit form of $f_h$ does not affect the powers of $(-k_i\eta_0)$ as in Eq.~\eqref{eq: int-scalar}, we will not provide it explicitly. A flattened configuration for the tensor perturbations gives $-c_h\tilde k_j\eta_0\ll1$.
While $\mathcal{I}_s$ shows a sharp peak at the flattened configuration, $\mathcal{I}_h$ does not since $\mathcal{I}_h=0$ for $\tilde k_j=0$. We treat $\mathcal{I}_h$ as a linear combination of two functions, both exhibiting sharp peaks at the flattened configuration, to evaluate the exponent of $(-k_i\eta_0)$.
We then decompose $\mathcal{I}_h$ as
\begin{align}
\mathcal{I}_h=\mathcal{I}_{h,1}+\mathcal{I}_{h,2},
\end{align}
where
\begin{align}
\mathcal{I}_{h,1}&:=if_h(k_i,s_i)\int\D\eta\frac{\D}{\D\eta}\biggl[(-\eta)^n e^{-ic_h\tilde k_j\eta}\biggr],\\
\mathcal{I}_{h,2}&:=inf_h(k_i,s_i)\int\D\eta(-\eta)^{n-1}e^{-ic_h\tilde k_j\eta}.
\end{align}
In the non-flattened configurations (where $|c_h\tilde k_j\eta|\gg1$ on the subhorizon scales), we have
\begin{align}
\mathcal{I}_{h}&\simeq\mathcal{I}_{h,2}\simeq in f_h(k_i,s_i)(-ic_h\tilde k_j)^{-n}\Gamma(n),\\
\mathcal{I}_{h,1}&\simeq0.
\end{align}
Note that the contour of this integration is actually displaced from the real axis, such as $\eta_0 \to \eta_0(1+i
\epsilon)$, due to the in-in formalism.
Conversely, in the flattened configurations (where $|c_h\tilde k_j\eta|\ll1$ on the subhorizon scales), we obtain
\begin{align}
\mathcal{I}_h&\simeq -\frac{c_h\tilde k_j\eta_0}{n+1}f_h(k_i,s_i)(-\eta_0)^n,\\
\mathcal{I}_{h,1}&\simeq f_h(k_i,s_i)(-\eta_0)^n(i-c_h\tilde k_j\eta_0),\\
\mathcal{I}_{h,2}&\simeq f_h(k_i,s_i)(-\eta_0)^n\biggl(-i+\frac{n}{n+1}c_h\tilde k_j\eta_0\biggr).
\end{align}
It is apparent that both $\mathcal{I}_{h,1}$ and $\mathcal{I}_{h,2}$ in the flattened configuration are substantially larger than their non-flattened counterparts, respectively. Therefore, $\mathcal{I}_{h,1}$ and $\mathcal{I}_{h,2}$ peak at the flattened configuration. The subsequent steps follow a similar process to that in the case of curvature perturbations.
The contributions from Eq.~(\ref{eq: int-tensor}) to the CMB bispectrum can be written as~\cite{Tahara:2017wud}
\begin{align}
\left\langle a^{(s_1)}_{l_1m_1}a^{(s_2)}_{l_2m_2}a^{(s_3)}_{l_3m_3}\right\rangle&\supset\sum_{i=1}^2\mathcal{F}_i,
\end{align}
where
\begin{widetext}
\begin{align}
\mathcal{F}_i&:=(4\pi)^3(-i)^{l_1+l_2+l_3}\int\frac{{\rm d}^3{\bf k}_1}{(2\pi)^3}\frac{{\rm d}^3{\bf k}_{2}}{(2\pi)^3}\frac{{\rm d}^3{\bf k}_{3}}{(2\pi)^3}\mathcal{T}^{(s_1)}_{l_1}(k_1)\mathcal{T}^{(s_{2})}_{l_{2}}(k_{2})\mathcal{T}^{(s_{3})}_{l_{3}}(k_{3})\notag\\
&\quad \times{}_{-s_1}Y^*_{l_1m_1}(\hat n_1){}_{-s_{2}}Y^*_{l_{2}m_{2}}(\hat n_{2}){}_{-s_{3}}Y^*_{l_{3}m_{3}}(\hat n_{3})(2\pi)^3\delta({\bf k}_1+{\bf k}_{2}+{\bf k}_{3})\mathcal{I}_{h,i}.
\end{align}
\end{widetext}
Here, $a^{(s)}_{lm}$ are the expansion coefficients of the CMB fluctuations in terms of the spin-weighted spherical harmonics and $\mathcal{T}^{(s)}_l(k)$ is the transfer function of the temperature fluctuation, E-mode polarization, or B-mode polarization originating from the tensor perturbations. 
After performing the ${\bf k}_{j+2}$-integral via the delta function and taking the flattened limit of $\mathcal{T}^{(s)}_l(k), {}_{-s}Y^*_{lm}(\hat n)$, and $f_h(k_i)$, we perform the $\hat n_j$-integral as
\begin{align}
\mathcal{F}_1&\propto\int\D^2\hat n_j \mathcal{I}_{h,1}(\tilde k_j)=2\pi\frac{k_j-k_{j+1}}{c_hk_jk_{j+1}}(-\eta_0)^{n-1}, \label{eq: ang-int-1}\\
\mathcal{F}_2&\propto\int\D^2\hat n_j \mathcal{I}_{h,2}(\tilde k_j)=\frac{2\pi n}{1-n}\frac{k_j-k_{j+1}}{c_hk_jk_{j+1}}(-\eta_0)^{n-1}, \label{eq: ang-int-2}
\end{align}
where we ignored rapidly oscillating terms such as $e^{i c_h k_{j+1} \eta}$ because they result in highly suppressed terms after $k_{j+1}$-integral.
For the flattened case ($|c_h\tilde k_j\eta_0|\ll1$), we have
\begin{align}
\mathcal{I}_h\propto(-\eta_0)^{n+1}, \label{eq: primordial-flat-eta}
\end{align}
indicating that the primordial bispectrum is proportional to $(-\eta_0)^{n+1}$, and thus two powers of $|c_hk\eta_0|$ are diminished in the CMB bispectra originating from the tensor modes. This is in contrast to the case of the scalar modes, where only one power of $(-\eta_0)$ is reduced. It should be noted here that the leading order contributions from both integrals in Eqs.~(\ref{eq: ang-int-1}) and~(\ref{eq: ang-int-2}) do not cancel out each other, i.e., $(\mathcal{F}_1+\mathcal{F}_2)\propto(-\eta_0)^{n-1}$. 
\footnotetext{The authors of Refs.~\cite{Kanno:2022mkx,Gong:2023kpe} obtained the three-point functions which do not vanish at the flattened limit. In using the in-in formalism, they first killed the contributions at $\eta=\eta_0$ to the three-point function for all of the triangles by taking $\eta_0\to-\infty(1+i\epsilon)$ and then took the explicit limits (e.g., the flattened limit) to the function obtained after the time integral. In this case, the resultant three-point function is singular at $\tilde k_j=0$. However, similarly to the calculations performed in the context of the scalar modes in Ref.~\cite{Holman:2007na}, the authors of Ref.~\cite{Akama:2020jko} kept $\eta_0$ finite and performed the time integrals separately for the flattened and non-flattened triangles of which integrand oscillates and does not oscillate at $\eta=\eta_0$, respectively. The three-point function obtained in this way is regular at $\tilde k_j=0$ and picks up the contributions at $\eta=\eta_0$ which are the consequence of interactions among the subhorizon modes whose physical momenta are $k/a(\eta_0)\sim\Lambda$.
This discrepancy comes from that Refs.~\cite{Holman:2007na,Akama:2020jko} count contributions from partial circular contour at large radius $(-\eta_0)$ of the in-in formalism, but Refs.~\cite{Kanno:2022mkx,Gong:2023kpe} do not.}

The enhancement was investigated within the GLPV theory in Ref.~\cite{Akama:2020jko}. The theory includes two tensor cubic operators in the form of $h^2\partial^2 h$ and $\dot h^3$. The former is present even in the Einstein-Hilbert action (i.e., in general relativity), while the latter is induced, e.g., in the Horndeski theory and in some classes beyond Horndeski such as the GLPV theory. In Ref.~\cite{Akama:2020jko}, it was found in the GLPV theory that the former and latter operators yield the $|k_i\eta_0|$-dependence on the bispectrum as $|k_i\eta_0|^2$ and $|k_i\eta_0|^3$, respectively. Given the previous argument on the reduction, only the operator $\dot h^3$ may retain the enhancement in the bispectrum within the GLPV theory. On the other hand, the effects of the non-Bunch-Davies tensor modes in the theories with only the cubic operator $h^2\partial^2 h$ are not enhanced in the CMB bispectra. In the following section, we first consider the GLPV theory and see whether the enhancements remain in the CMB bispectra. We also explore the potential to attain greater enhancements in non-Gaussianities than those within the GLPV theory.

\section{Possible enhancements of CMB bispectra}\label{sec: CMB-bispectra-GLPV-beyond}
In this section, we investigate a potential for non-Bunch-Davies effects to enhance CMB bispectra within the GLPV and beyond-GLPV theories.
To do this, we introduce the following dimensionless parameter:
\begin{align}
f^{\rm CMB}_{\rm NL}:=f_{\rm NL}\biggl(\frac{c_h\Lambda}{H}\biggr)^{-2},
\end{align}
where
\begin{align}
f_{\rm NL}:=\frac{\mathcal{B}_h}{(\mathcal{P}_h^*)^2}\frac{k_1^3k_2^3k_3^3}{\sum_i k_i^3}, \label{eq: fnl}
\end{align}
with $\mathcal{P}_h^*$ being the dimensionless tensor power spectrum $\mathcal{P}_h$ evaluated at the end of inflation, and $\mathcal{P}_h$ is defined by
\begin{align}
\langle h_{ij}({\bf k})h_{ij}({\bf k}')\rangle=(2\pi)^3\delta({\bf k}+{\bf k}')\frac{2\pi^2}{k^3}\mathcal{P}_h.
\end{align}
In our setup, the power spectrum at the end of inflation reads~\cite{Akama:2020jko}
\begin{align}
\mathcal{P}_h^*=\frac{1}{\pi^2}\frac{H^2}{c_h\mathcal{F}_T}\sum_s\left|\alpha^{(s)}_k-\beta^{(s)}_k\right|^2=\mathcal{O}\biggl(\frac{H^2}{c_h\mathcal{F}_T}\biggr),
\end{align}
where we have assumed $|\alpha^{(s)}_k|, |\beta^{(s)}_k|\lesssim\mathcal{O}(1)$ since both Bogoliubov coefficients satisfy the normalization condition, Eq.~(\ref{eq: normali-bogo}), and the backreaction constraint indicates $|\beta^{(s)}_k|\lesssim\mathcal{O}(1)$ which will be shown later. Eq.~(\ref{eq: fnl}) is analogous to the conventional non-linearity parameter for the scalar non-Gaussianity. The factor $(c_h\Lambda/H)^{-2}$ is required to discuss the amplitude relevant to the CMB bispectra (i.e., to take into account the reduction of two powers of $|k_i\eta_0|$). In the following subsections, we investigate whether $f^{\rm CMB}_{\rm NL}$ can be enhanced due to the non-Bunch-Davies effects or not.

Hereafter, we consider both GLPV and beyond-GLPV theories in the ADM formalism as described in several studies~\cite{Gleyzes:2014dya,Gleyzes:2014qga,Gao:2014soa}.

\subsection{GLPV}
The ADM Lagrangian of the GLPV theory is of the form~\cite{Gleyzes:2014dya,Gleyzes:2014qga}:
\begin{align}
\mathcal{L}_{\rm GLPV}&=A_2(t,N)+A_3(t,N)K+A_4(t,N)\left(K^2-K_{ij}^2\right)
\notag\\
&\ +B_4R+A_5\left(K^3-3KK_{ij}^2+2K_{ij}^3\right)\notag\\
&\ +B_5\left(K_j^iR_i^j-\frac{1}{2}KR\right), \label{eq: lag-GLPV}
\end{align}
where $A_i\ (i=2,3,4,5)$ and $B_i\ (i=4,5)$ are arbitrary functions of $t$ and $N$, $K_{ij}$ and $R_{ij}$ are the extrinsic and intrinsic curvature tensors, respectively, defined on $t$-{\rm constant} hypersurfaces, and $K:=\gamma^{ij}K_{ij}$ and $R:=\gamma^{ij}R_{ij}$ are their traces. In particular, the above Lagrangian with constraints $A_4=-B_4-N\partial B_5/\partial N$ and $A_5=(N/6)\partial B_5/\partial N$ reproduces the Lagrangian of the Horndeski theory. 
Eq.~\eqref{eq: lag-GLPV} is written as a spatially covariant Lagrangian respecting only three-dimensional covariance, but the four-dimensional covariance can be restored using the St\"{u}ckelberg trick. (See, e.g., Ref.~\cite{Gleyzes:2014dya} for the GLPV Lagrangian respecting the four-dimensional covariance.)

In this theory, the quadratic action takes the form of Eq.~(\ref{eq: quad-tensor}) with
\begin{align}
\mathcal{G}_T&=-2(A_4+3A_5H),\\
\mathcal{F}_T&=2B_4+\dot B_5,
\end{align}
where a dot denotes differentiation with respect to $t$, and the cubic Lagrangian is of the form
\begin{align}
\mathcal{L}^{(3)}_{h,\rm GLPV}&=\frac{\mathcal{F}_T}{4a^2}\biggl(h_{ik}h_{jl}-\frac{1}{2}h_{ij}h_{kl}\biggr)\partial_k\partial_l h_{ij}\notag\\
&\quad +\frac{A_5}{4}\dot h_{ij}\dot h_{jk}\dot h_{ki}. \label{eq: cubic-GLPV}
\end{align}
Here, we have assumed $\mathcal{G}_T, \mathcal{F}_T={\rm const.}$ in the de Sitter background, which means $A_4, A_5, B_4, \dot B_5={\rm const.}$
In this theory, the primordial bispectrum of the tensor perturbations has been obtained in Ref.~\cite{Akama:2020jko}. In particular, the explicit form for the nearly-flattened triangle up to the leading-order in $\beta^{(s)}_k$ reads\footnote{The bispectrum includes the terms of higher-order in $\beta^{(s)}_k$, but those terms are at most the same magnitudes with Eqs.~(\ref{eq: B1-GLPV}) and~(\ref{eq: B2-GLPV}) when $\beta^{(s)}_k$ takes the maximum value which is ${O}(1)$ in the present paper. Therefore, Eqs.~(\ref{eq: B1-GLPV}) and~(\ref{eq: B2-GLPV}) are sufficient to consider when we estimate the amplitude of the bispectrum.}
\begin{align}
\mathcal{B}_h=\mathcal{B}_{\mathcal{F}_T}+\mathcal{B}_{A_5},
\end{align}
where
\begin{align}
\mathcal{B}_{\mathcal{F}_T}&\simeq\frac{2H^4}{c_h^2\mathcal{F}_T^2}\frac{1}{k_1^3k_2^3k_3^3}(s_1k_1+s_2k_2+s_3k_3)^2F(s_i,k_i)\notag\\
&\quad\times\biggl[\mathcal{I}_0(k_1,k_2,k_3)-\frac{k_1k_2k_3}{2}c_h^2\eta_0^2{\rm Re}[\beta^{(s_1)}_{k_1}]\biggr], \label{eq: B1-GLPV}\\
\mathcal{B}_{A_5}&\simeq\frac{192A_5H^5}{\mathcal{F}_T^3}\frac{F(s_i,k_i)}{k_1k_2k_3}\biggl[\frac{1}{K^3}-\frac{c_h^3\eta_0^3}{6}{\rm Im}[\beta^{(s_1)}_{k_1}]\biggr], \label{eq: B2-GLPV}
\end{align}
with $\mathcal{B}_{\mathcal{F}_T}$ and $\mathcal{B}_{A_5}$ being the bispectrum originating from the first and second cubic operators in Eq.~(\ref{eq: cubic-GLPV}), respectively. We also defined
\begin{align}
\mathcal{I}_0(k_1,k_2,k_3)&:=-K+\frac{k_1k_2k_3}{K^2}+\frac{k_1k_2+k_2k_3+k_3k_1}{K},\\
F(s_i,k_i)&:=\frac{1}{64}\frac{K}{k_1^2k_2^2k_3^2}(s_1k_1+s_2k_2+s_3k_3)^2\notag\\
&\quad\times(k_1-k_2-k_3)(k_1-k_2+k_3)\notag\\
&\quad\times(k_1+k_2-k_3),
\end{align}
with $K:=k_1+k_2+k_3$. The $k_i\eta_0$-dependent terms in Eqs.~\eqref{eq: B1-GLPV} and~\eqref{eq: B2-GLPV} are obtained from the time integral~\eqref{eq: int-tensor} with $n=1$ and $n=2$, respectively. 
Considering our previous argument on the reduction of $|c_hk_i\eta_0|^2$, the non-linearity parameter $f^{\rm CMB}_{\rm NL}$ derived from $\mathcal{B}_{\mathcal{F}_T}$ is no longer enhanced, and thus we focus solely on $\mathcal{B}_{A_5}$.  The magnitude of the non-linearity parameter contributed from the excited modes (i.e., $\beta^{(s)}_k$ terms) can be computed as
\begin{align}
f^{\rm CMB}_{\rm NL}=\mathcal{O}\biggl(\frac{A_5 H}{\mathcal{G}_T}|\beta^{(s_1)}_{k_1}|(-c_hk_i\eta_0)\biggr).
\end{align}
To discuss the potential enhancements in the observable quantities, we investigate an upper bound on $f^{\rm CMB}_{\rm NL}$. To do so, we consider theoretical constraints on the magnitude of $\beta^{(s)}_k$ and that of the coupling function of the cubic operator. The first constraint comes from the argument on the backreaction from the excited modes. The modes which get excited from the Bunch-Davies state cause backreaction to the inflationary background~\cite{Tanaka:2000jw}. When the quadratic action is of the form Eq.~(\ref{eq: quad-tensor}), the condition to prevent the excited modes from disrupting the inflationary background has been obtained as~\cite{Akama:2020jko}
\begin{align}
\frac{c_h}{a^4(\eta_0)}\int^{\Lambda a(\eta_0)}|\beta^{(s)}_k|^2k^3\D k\lesssim \mpl^2 H^2. \label{eq: back-generic}
\end{align}
For simplicity, we assume 
\begin{align}
\beta^{(s)}_k\sim
\begin{cases}
\beta & {\rm for}\ k\leq \Lambda a(\eta_0) \\
0 & {\rm for}\ k> \Lambda a(\eta_0),
\end{cases}
\end{align}
where $\beta$ is constant. Hence, Eq.~(\ref{eq: back-generic}) can be rewritten as\footnote{As has been shown in Ref.~\cite{Akama:2020jko}, the same form can be obtained from the following ansatz:
\begin{align}
\beta^{(s)}_k\sim\beta\exp[-k^2/(\Lambda a(\eta_0))^2].
\end{align}}
\begin{align}
|\beta|\lesssim(\mathcal{P}_h^*)^{1/2}\frac{\mpl}{\Lambda}\frac{\mathcal{F}_T^{1/2}}{\Lambda}. \label{eq: back-const}
\end{align}
The second constraint can be derived from the following perturbativity condition:
\begin{align}
\mathcal{L}^{(2)}_h>\mathcal{L}^{(3)}_h, \label{eq: cond-perturbation}
\end{align}
where $\mathcal{L}^{(2)}_h$ and $\mathcal{L}^{(3)}_h$ are the quadratic and cubic Lagrangians of the tensor perturbations, respectively. Eq.~\eqref{eq: cond-perturbation} is necessary as long as the solution of the linear perturbation is used. See also Ref.~\cite{Leblond:2008gg} for a similar perturbativity argument.
We evaluate Eq.~(\ref{eq: cond-perturbation}) at $\eta=\eta_0$ in the GLPV theory and get
\begin{align}
\frac{A_5 H}{\mathcal{G}_T}<\frac{H}{c_h\Lambda}|h_{ij}|^{-1}|_{\eta=\eta_0}. \label{eq: const-sc}
\end{align}
We estimate the amplitude of $h_{ij}$ from the primordial power spectrum as\footnote{More specifically, we ignored the $\log(k_{\rm UV}/k_{\rm IR})$-term compared to the $|c_hk_{\rm UV}\eta_0|^2$-term where $k_{\rm UV}=a(\eta_0)\Lambda$ stands for the UV cutoff and $k_{\rm IR}$ does the IR one.}
\begin{align}
\mathcal{O}(|h_{ij}|^2)|_{\eta=\eta_0}=\int^{a(\eta_0)\Lambda}\frac{\D k}{k}\mathcal{P}_h|_{\eta=\eta_0}=\mathcal{O}\biggl(\mathcal{P}^*_{h}\frac{c_h^2\Lambda^2}{H^2}\biggr), \label{eq: h-amp}
\end{align}
where we used the following form of the power spectrum
\begin{align}
\mathcal{P}_h|_{\eta=\eta_0}=\mathcal{P}^*_h(1+c_h^2k^2\eta_0^2)\simeq \mathcal{P}^*_hc_h^2k^2\eta_0^2.
\end{align}
Note that we can treat $h_{ij}$ as perturbations at $\eta=\eta_0$ under the condition:
\begin{align}
\Lambda^2\ll \sqrt{\mathcal{G}_T\mathcal{F}_T}.
\label{eq: cond-cut}
\end{align}
This condition stems from the requirement that $|h_{ij}|\ll1$. Then, by combining Eqs.~(\ref{eq: back-const}),~(\ref{eq: const-sc}) and~(\ref{eq: h-amp}), one can evaluate the upper bound on $f^{\rm CMB}_{\rm NL}$ as
\begin{align}
f^{\rm CMB}_{\rm NL}\lesssim|\beta||h_{ij}|^{-1}\lesssim\frac{\mpl}{\Lambda}\frac{\mathcal{F}_T^{1/2}}{\Lambda}\frac{H}{c_h\Lambda}.
\end{align}
Assuming $\mathcal{G}_T, \mathcal{F}_T\sim\mpl^2$ as typical values, we find
\begin{align}
f^{\rm CMB}_{\rm NL}\lesssim \frac{H}{\Lambda}\biggl(\frac{\mpl}{\Lambda}\biggr)^2. \label{eq: const-fNL-GLPV}
\end{align}
In this case, Eq.~(\ref{eq: cond-cut}) indicates $\Lambda\ll\mpl$. For a cutoff scale enjoying $H<\Lambda\ll\mpl$, the resultant parameter $f^{\rm CMB}_{\rm NL}$ can relatively be amplified. Under our setup, the perturbations are on the subhorizon scales at $\eta=\eta_0$, and thus we take $\Lambda\sim10^2 H$ (which implies $|k_i\eta_0|\lesssim\mathcal{O}(10^2)$) as a possible lowest cutoff scale. Then we find 
\begin{align}
f^{\rm CMB}_{\rm NL}\lesssim\mathcal{O}(10^5),
\end{align}
where we have assumed $H^2/\mpl^2\lesssim\mathcal{O}(10^{-10})$ in accordance with the current constraint on the tensor-to-scalar ratio, $r\lesssim \mathcal{O}(10^{-2})$~\cite{Planck:2018jri}. Note that $\beta^{(s)}_k=\mathcal{O}(1)$ in the case of the possible lowest cutoff scale.

Here, the primordial bispectrum explicitly depends on $\eta_0$, which implies that the flattened non-Gaussianity is generated on the subhorizon scales. This might lead one to expect that a higher-derivative cubic operator could yield a larger $f^{\rm CMB}_{\rm NL}$. In the following subsection, we investigate whether $f^{\rm CMB}_{\rm NL}$ is further enhanced in an extended gravitational theory yielding higher-derivative cubic operators.

\subsection{Beyond GLPV}
Let us consider the following Lagrangian:
\begin{align}
\mathcal{L}=\mathcal{L}_{\rm GLPV}+\mathcal{L}_{\rm ex}, \label{eq: Lag-GLPV-ex}
\end{align}
where
\begin{align}
\mathcal{L}_{\rm ex}&=C_1K_{ik}K_{kj}R^{(3)}_{ij}+C_2\biggl[-\frac{1}{3}K(R^{(3)}_{ij})^2+K^i_j R^{(3)}_{ki}R^{(3)}_{kj}\biggr]\notag\\
&\quad +C_3(R^{(3)}_{ij})^3, \label{eq: ex-L}
\end{align}
where $C_i(i=1,2,3)$ are the arbitrary functions of $t$ and $N$, and we assume that $C_i$'s are almost constant in de Sitter background. Those terms are a subclass of the general spatially-covariant theory beyond GLPV~\cite{Gao:2014soa}.

A property of this subclass is that the quadratic action of the tensor perturbations is of the standard form Eq.~\eqref{eq: quad-tensor}, while the cubic Lagrangian includes terms with higher derivatives than those in the GLPV theory\footnote{The Lagrangian proposed in Ref.~\cite{Gao:2014soa} includes the GLPV term with arbitrary coefficients, e.g., $\mathcal{L}\supset A_4(t,N)K^2, \tilde A_4(t,N)K_{ij}^2$, where the both coefficients are independent of each other. Since such GLPV terms do not yield higher-derivative cubic operators, we do not consider them in the present paper.}:
\begin{align}
\mathcal{L}^{(3)}_{h,\rm beyond}&=\mathcal{L}^{(3)}_{h,\rm GLPV}+\mathcal{L}^{(3)}_{h,\rm ex},
\end{align}
where
\begin{align}
\mathcal{L}^{(3)}_{h,\rm ex}&=a^3\biggl[-\frac{C_1}{8a^2}\dot h^i_k\dot h^k_j\partial^2 h^j_i+\frac{C_2}{8a^4}\dot h^i_k\partial^2 h^k_j\partial^2 h^j_i\notag\\
&\quad\ -\frac{C_3}{8a^6}(\partial^2 h_{ij})^3\biggr]. \label{eq: cubic-XG3}
\end{align}
In the general class of the spatially-covariant theory, the quadratic action is modified by the $(\partial^2 h_{ij})^2$ term~\cite{Ashoorioon:2011eg,Akita:2015mho}. Since our purpose here is to investigate the enhancements from higher-derivative cubic operators, the extra Lagrangian given by Eq.~(\ref{eq: ex-L}) is sufficient for this purpose.

Three-point correlation functions can be calculated straightforwardly, and we leave the details of the calculations to Appendix~\ref{sec: App}. The $\beta^{(s)}_k$ terms in the primordial bispectra from the extra cubic operators take the following forms:
\begin{align}
\mathcal{B}_{C_1}&\sim\frac{C_1 H^6}{c_h^2\mathcal{F}_T^3}\frac{1}{k_i^6}\beta^{(s)}_k|c_hk_i\eta_0|^3,\\
\mathcal{B}_{C_2}&\sim \frac{C_2H^7}{c_h^4\mathcal{F}_T^3}\frac{1}{k_i^6}\beta^{(s)}_k|c_hk_i\eta_0|^5,\\
\mathcal{B}_{C_3}&\sim \frac{C_3H^8}{c_h^6\mathcal{F}_T^3}\frac{1}{k_i^6}\beta^{(s)}_k|c_hk_i\eta_0|^6,
\end{align}
where the subscript in $\mathcal{B}_\bullet$ denotes which term the bispectrum arises from. 
From Eq.~(\ref{eq: cond-perturbation}), we have
\begin{align}
C_1&<\frac{\mathcal{G}_T}{\Lambda^2}|h_{ij}|^{-1},\label{eq: const-b6}\\
C_2&<\frac{\mathcal{F}_T}{c_h}\frac{1}{\Lambda^3}|h_{ij}|^{-1}, \label{eq: const-a7}\\
C_3&<\frac{\mathcal{F}_T}{\Lambda^4}|h_{ij}|^{-1}. \label{eq: const-d7}
\end{align}
The explicit forms of $\mathcal{G}_T, \mathcal{F}_T$, and $c_h^2$ are different between the GLPV theory and the beyond GLPV theory, 
\begin{align}
\mathcal{G}_T&=-2(A_4+3A_5H),\\
\mathcal{F}_T&=2B_4+\dot B_5+3C_1 H^2+2\frac{\D}{\D t}(C_1H).
\end{align}
In Ref.~\cite{Akama:2020jko}, the backreaction constraint was obtained only within the GLPV theory. Since the quadratic action of both the GLPV and beyond-GLPV theory has the same form as Eq.~\eqref{eq: quad-tensor}, we can use Eq.~(\ref{eq: back-const}) in the beyond-GLPV theory as well. Finally, combining Eqs.~(\ref{eq: back-const}), (\ref{eq: const-b6}), (\ref{eq: const-a7}), and (\ref{eq: const-d7}), we derive
\begin{align}
f^{\rm CMB}_{{\rm NL},{C_1}}&\lesssim \frac{\mpl}{\Lambda}\frac{\mathcal{F}_T^{1/2}}{\Lambda}\biggl(\frac{H}{c_h\Lambda}\biggr)^2,\\
f^{\rm CMB}_{{\rm NL},{(C_2,C_3)}}&\lesssim \frac{\mpl}{\Lambda}\frac{\mathcal{F}_T^{1/2}}{\Lambda}\frac{H}{c_h\Lambda}, \label{eq: const-fNL-a7}
\end{align}
where $f^{\rm CMB}_{{\rm NL},\bullet}$ stands for $f^{\rm CMB}_{{\rm NL}}$ originating from $\mathcal{B}_{\bullet}$.
The requirement for the perturbation to be on the subhorizon scales at $\eta=\eta_0$ is $|c_hk_i\eta_0|\gg1$, implying $c_h\Lambda/H\gg1$. Thus, the more stringent condition on $f^{\rm CMB}_{\rm NL}$ is obtained from Eq.~(\ref{eq: const-fNL-a7}). It should be emphasized here that Eq.~(\ref{eq: const-fNL-a7}) is exactly the same as Eq.~(\ref{eq: const-fNL-GLPV}). Therefore, though the resultant $f^{\rm CMB}_{\rm NL}$ can indeed be amplified when the cutoff scale is close to $H$, one cannot easily enhance $f^{\rm CMB}_{\rm NL}$ even by introducing higher-derivative cubic operators in extended theories of gravity because of the perturbativity condition Eq.~(\ref{eq: cond-perturbation}).

Before concluding this section, it is noteworthy to highlight a potential advantage offered by the enhancement in the flattened limit. In gravity theories devoid of parity violation, the B-mode auto-bispectrum vanishes under the geometrical condition of $l_i = l_j~(i \ne j)$. The primordial bispectrum enhanced around the flattened configuration ($k_1=k_2+k_3$) implies that the CMB bispectrum would also be enhanced around $l_1\simeq l_2+l_3$, which does not conflict the condition above.
Conversely, the primordial bispectrum amplified around the squeezed ($k_1\simeq 0$) or equilateral ($k_1 \simeq k_2 \simeq k_3$) configuration results in the CMB bispectrum peaking around $l_1\simeq 0$ or $l_1 \simeq l_2 \simeq l_3$, respectively, either of which are suppressed due to the aforementioned geometrical condition. Hence, when observing the B-mode bispectrum, one could anticipate that the enhancement of the primordial bispectrum in the flattened limit would exhibit a relative advantage over that in the squeezed or equilateral limit.

\section{Summary}\label{sec: summary}
In the present paper, we first clarified that the $(-k_i\eta_0)^n$-dependence in the primordial tensor bispectrum yields $(-k_i\eta_0)^{n-2}$-enhancement in the CMB bispectra. We then found that the $(-k_i\eta_0)^n$-dependence obtained from the cubic operators present in the Einstein-Hilbert action does not lead to any enhancements in the CMB bispectra. We also showed that the CMB bispectra can enhance in extended gravitational theories, the GLPV theory and its extensions. In the case of the Bunch-Davies states, the primordial tensor auto-bispectrum for the exact-flattened triangle ($\tilde k_j=0$) vanishes and that for the nearly-flattened one is not enhanced but just suppressed in proportion to $\tilde k_j$. Therefore, our results indicate that any detection of the tensor flattened non-Gaussianities by CMB experiments would support inflation models with non-Bunch-Davies states in such extended theories of gravity involving higher-derivative cubic operators.

In evaluating enhancement, we introduced a dimensionless quantity $f^{\rm CMB}_{\rm NL}$ and derived its upper bound which is determined from the backreaction constraint and the perturbativity condition. Our analysis indicates that cubic operators involving higher-order derivatives do not necessarily lead to a larger $f^{\rm CMB}_{\rm NL}$. This is due to the fact that higher-derivative terms are significantly constrained by the perturbativity condition. 
It would be interesting to look for extended theories of gravity that can have a more impact on the CMB bispectra.

As a further study, it would also be important to compute the CMB bispectra numerically and evaluate the signal-to-noise ratio. The enhancement around the flattened triangle occurs only for a very limited angle, and as estimated in Ref.~\cite{Holman:2007na}, some of the signals could be buried in noise. The detailed analysis is beyond the scope of this paper, and we will leave it for future work.

\section*{Acknowledgments}
We would like to thank Tsutomu Kobayashi for collaboration in the early stage of this work and for useful comments on the manuscript. We would like to thank Paola C. M. Delgado, Shin'ichi Hirano, Chunshan Lin, Giorgio Orlando, and Shuichiro Yokoyama for helpful discussions. We would like to thank Jinn-Ouk Gong, Maria Mylova, and Misao Sasaki for helpful correspondence.
SA was supported by the grants No. UMO-2018/30/Q/ST9/00795 and No. UMO-2021/42/E/ST9/00260 from the National Science Centre, Poland. 
HWHT was supported by Grant-in-Aid for JSPS Fellows 21F21019.
\begin{widetext}
\appendix
\section{Bispectra from Extra Terms} \label{sec: App}

By using the in-in formalism, one can calculate the three-point function of the tensor perturbations as
\begin{align}
\langle\xi^{(s_1)}({\bf k}_1)\xi^{(s_2)}({\bf k}_2)\xi^{(s_3)}({\bf k}_3)\rangle=-i\int_{\eta_0}^0\D\eta a(\eta)\langle[\xi^{(s_1)}(0,{\bf k}_1)\xi^{(s_2)}(0,{\bf k}_2)\xi^{(s_3)}(0,{\bf k}_3),H_{\rm int}(\eta)]\rangle,
\end{align}
where the interaction Hamiltonian $H_{\rm int}$ is defined by
\begin{align}
H_{\rm int}:=-\int\D^3x\mathcal{L}^{(3)}_h,
\end{align}
with $\mathcal{L}^{(3)}_h$ being the cubic Lagrangian of the tensor perturbations. The primordial bispectrum in the GLPV theory has been calculated in Ref.~\cite{Akama:2020jko}, and thus we here show the results only from the extra terms in the beyond-GLPV theory. For convenience, we define the resultant bispectrum as
\begin{align}
\mathcal{B}_\bullet={\rm Re}[\tilde{\mathcal{B}}_\bullet].
\end{align}
First, we compute the bispectrum originating from the $C_1$ term. By employing the in-in formalism, one can write the bispectrum as
\begin{align}
\mathcal{B}_{C_1}={\rm Re}[\tilde{\mathcal{B}}_{C_1}],
\end{align}
where
\begin{align}
\tilde{\mathcal{B}}_{\rm C_1}&=-i\frac{4C_1c_hH^6}{\mathcal{F}_T^3}\frac{1}{k_1k_2k_3}\Pi_i\left(\alpha_{k_i}^{(s_i)}-\beta_{k_i}^{(s_i)}\right)\biggl[\alpha_{k_1}^{(s_1)*}\alpha_{k_2}^{(s_2)*}\alpha_{k_3}^{(s_3)*}I_{\rm {C_1},1}+\beta_{k_1}^{(s_1)*}\beta_{k_2}^{(s_2)*}\beta_{k_3}^{(s_3)*}I_{\rm {C_1},2}\notag\\
&\quad +\biggl(\alpha_{k_1}^{(s_1)*}\alpha_{k_2}^{(s_2)*}\beta_{k_3}^{(s_3)*}I_{\rm {C_1},3}+(k_1,s_1\leftrightarrow k_2,s_2)+(k_1,s_1\leftrightarrow k_3,s_3)\biggr)\notag\\
&\quad +\biggl(\beta_{k_1}^{(s_1)*}\beta_{k_2}^{(s_2)*}\alpha_{k_3}^{(s_3)*}I_{\rm {C_1},4}+(k_1,s_1\leftrightarrow k_2,s_2)+(k_1,s_1\leftrightarrow k_3,s_3)\biggr)\biggr]F(s_i,k_i),
\end{align}
with
\begin{align}
I_{\rm {C_1},1}&=\int\D\eta\eta^2(-3+ic_hK\eta)e^{ic_hK\eta},\\
I_{\rm {C_1},2}&=\int\D\eta\eta^2(3+ic_hK\eta)e^{-ic_hK\eta},\\
I_{\rm {C_1},3}&=\int\D\eta\eta^2(3-ic_h\tilde k\eta)e^{ic_h\tilde k\eta},\\
I_{\rm {C_1},4}&=\int\D\eta\eta^2(-3-ic_h\tilde k\eta)e^{-ic_h\tilde k\eta},
\end{align}
where $\tilde k:=-k_1+k_2+k_3$. First, we consider the non-flattened limit enjoying $|c_h\tilde k\eta_0|\gg1$. In this limit, we have
\begin{align}
{\rm Re}[I_{C_1,i}]&=0,\\
{\rm Im}[I_{C_1,1}]&={\rm Im}[I_{C_1,2}]=-\frac{12}{c_h^3K^3},\\
{\rm Im}[I_{C_1,3}]&={\rm Im}[I_{C_1,4}]=\frac{12}{c_h^3\tilde k^3},
\end{align}
where $K:=k_1+k_2+k_3$. 
Finally, we obtain
\begin{align}
\tilde{\mathcal{B}}_{\rm C_1}&=-\frac{48C_1H^6}{c_h^2\mathcal{F}_T^3}\frac{1}{k_1k_2k_3}\biggl[\Pi_i(\alpha^{(s_i)}_{k_i}-\beta^{(s_i)}_{k_i})\biggr]\biggl\{\biggl(\alpha^{(s_1)*}_{k_1}\alpha^{(s_2)*}_{k_2}\alpha^{(s_3)*}_{k_3}+\beta^{(s_1)*}_{k_1}\beta^{(s_2)*}_{k_2}\beta^{(s_3)*}_{k_3}\biggr)\frac{1}{K^3}\notag\\
&\quad -\biggl[\biggl(\alpha_{k_1}^{(s_1)*}\alpha_{k_2}^{(s_2)*}\beta_{k_3}^{(s_3)*}+\beta_{k_1}^{(s_1)*}\beta_{k_2}^{(s_2)*}\alpha_{k_3}^{(s_3)*}\biggr)\frac{1}{\tilde k^3}+(k_1,s_1\leftrightarrow k_2,s_2)+(k_1,s_1\leftrightarrow k_3,s_3)\biggr]\biggr\}.
\end{align}
Then, we consider the flattened limit enjoying $|c_h\tilde k\eta_0|\ll1$. In this case, we obtain
\begin{align}
{\rm Re}[I_{C_1,1}]&={\rm Re}[I_{C_1,2}]=0,\\
{\rm Re}[I_{C_1,3}]&=-{\rm Re}[I_{C_1,4}]=-\eta_0^3,\\
{\rm Im}[I_{C_1,1}]&={\rm Im}[I_{C_1,2}]=-\frac{12}{c_h^3K^3},\\
{\rm Im}[I_{C_1,3}]&={\rm Im}[I_{C_1,4}]=-\frac{1}{2}c_h\tilde k\eta_0^4,
\end{align}
and hence we have
\begin{align}
\tilde{\mathcal{B}}_{\rm C_1}&=-\frac{48C_1H^6}{c_h^2\mathcal{F}_T^3}\biggl\{\frac{1}{k_1k_2k_3}\biggl[\Pi_i(\alpha^{(s_i)}_{k_i}-\beta^{(s_i)}_{k_i})\biggr]\biggl[\biggl(\alpha^{(s_1)*}_{k_1}\alpha^{(s_2)*}_{k_2}\alpha^{(s_3)*}_{k_3}+\beta^{(s_1)*}_{k_1}\beta^{(s_2)*}_{k_2}\beta^{(s_3)*}_{k_3}\biggr)\frac{1}{K^3}\notag\\
&\quad -\frac{i}{12}c_h^3\eta_0^3\biggl(\alpha_{k_1}^{(s_1)*}\alpha_{k_2}^{(s_2)*}\beta_{k_3}^{(s_3)*}-\beta_{k_1}^{(s_1)*}\beta_{k_2}^{(s_2)*}\alpha_{k_3}^{(s_3)*}\biggr)\biggr]F(s_i,k_i)\biggr\}\biggl|_{\tilde k\to0},
\end{align}
where we used 
\begin{align}
{\rm Re}[I_{\rm C_1,{(3,4)}}]\gg{\rm Im}[I_{\rm C_1,{(3,4)}}].
\end{align}
The integral which characterizes the $\eta_0$-dependence of the bispectrum is
\begin{align}
\int\D\eta\eta^2 e^{ic_h\tilde k\eta}.
\end{align}
One can compute the bispectra from the other two terms similarly, and thus we show only the results below. Regarding the $C_2$ term, the bispectrum evaluated at the non-flattened and flattened limits are obtained, respectively, as
\begin{align}
\tilde{\mathcal{B}}_{C_2}&=\frac{24C_2H^7}{c_h^4\mathcal{F}_T^3}\frac{1}{k_1k_2k_3}\Pi_i\left(\alpha_{k_i}^{(s_i)}-\beta_{k_i}^{(s_i)}\right)\biggl\{\biggl(\alpha_{k_1}^{(s_1)*}\alpha_{k_2}^{(s_2)*}\alpha_{k_3}^{(s_3)*}+\beta_{k_1}^{(s_1)*}\beta_{k_2}^{(s_2)*}\beta_{k_3}^{(s_3)*}\biggr)\frac{1}{K^3}\biggl(3+4\frac{k_1k_2+k_2k_3+k_1k_3}{K^2}\biggr)\notag\\
&\quad -\biggl[\biggl(\alpha_{k_1}^{(s_1)*}\alpha_{k_2}^{(s_2)*}\beta_{k_3}^{(s_3)*}+\beta_{k_1}^{(s_1)*}\beta_{k_2}^{(s_2)*}\alpha_{k_3}^{(s_3)*}\biggr)\frac{1}{\tilde k^3}\biggl(3+4\frac{k_1k_2-k_2k_3-k_1k_3}{\tilde k^2}\biggr)+(k_1,s_1\leftrightarrow k_2,s_2)+(k_1,s_1\leftrightarrow k_3,s_3)\biggr]\biggr\}\notag\\
&\quad\times F(s_i,k_i),
\end{align}
and
\begin{align}
\tilde{\mathcal{B}}_{C_2}&=\frac{24C_2H^7}{c_h^4\mathcal{F}_T^3}\biggl\{\frac{1}{k_1k_2k_3}\Pi_i\left(\alpha_{k_i}^{(s_i)}-\beta_{k_i}^{(s_i)}\right)\biggl[\biggl(\alpha_{k_1}^{(s_1)*}\alpha_{k_2}^{(s_2)*}\alpha_{k_3}^{(s_3)*}+\beta_{k_1}^{(s_1)*}\beta_{k_2}^{(s_2)*}\beta_{k_3}^{(s_3)*}\biggr)\frac{1}{K^3}\biggl(3+4\frac{k_1k_2+k_2k_3+k_1k_3}{K^2}\biggr)\notag\\
&\quad -\frac{i}{30}c_h^5(k_1^2+k_1k_2+k_2^2)\eta_0^5\biggl(\alpha_{k_1}^{(s_1)*}\alpha_{k_2}^{(s_2)*}\beta_{k_3}^{(s_3)*}-\beta_{k_1}^{(s_1)*}\beta_{k_2}^{(s_2)*}\alpha_{k_3}^{(s_3)*}\biggr)\biggr]F(s_i,k_i)\biggr\}\biggr|_{\tilde k\to0}.
\end{align}
The integral which defines the $\eta_0$-dependence of the bispectrum is
\begin{align}
\int\D\eta\eta^4 e^{ic_h\tilde k\eta}.
\end{align}
Regarding the $C_3$ term, the bispectrum evaluated at the non-flattened and flattened limits are obtained, respectively, as
\begin{align}
\tilde{\mathcal{B}}_{C_3}&=-\frac{96C_3H^8}{c_h^6\mathcal{F}_T^3}\frac{1}{k_1k_2k_3}\Pi_i\left(\alpha_{k_i}^{(s_i)}-\beta_{k_i}^{(s_i)}\right)\notag\\
&\quad \times\biggl\{\biggl(\alpha_{k_1}^{(s_1)*}\alpha_{k_2}^{(s_2)*}\alpha_{k_3}^{(s_3)*}+\beta_{k_1}^{(s_1)*}\beta_{k_2}^{(s_2)*}\beta_{k_3}^{(s_3)*}\biggr)\frac{1}{K^3}\biggl(1+3\frac{k_1k_2+k_2k_3+k_1k_3}{K^2}+15\frac{k_1k_2k_3}{K^3}\biggr)\notag\\
&\quad -\biggl[\biggl(\alpha_{k_1}^{(s_1)*}\alpha_{k_2}^{(s_2)*}\beta_{k_3}^{(s_3)*}+\beta_{k_1}^{(s_1)*}\beta_{k_2}^{(s_2)*}\alpha_{k_3}^{(s_3)*}\biggr)\frac{1}{\tilde k^3}\biggl(1+3\frac{k_1k_2-k_2k_3-k_1k_3}{\tilde k^2}-15\frac{k_1k_2k_3}{\tilde k^3}\biggr)\notag\\
&\quad\quad +(k_1,s_1\leftrightarrow k_2,s_2)+(k_1,s_1\leftrightarrow k_3,s_3)\biggr]\biggr\}F(s_i,k_i),
\end{align}
and
\begin{align}
\tilde{\mathcal{B}}_{C_3}&=-\frac{96C_3H^8}{c_h^6\mathcal{F}_T^3}\biggl\{\frac{1}{k_1k_2k_3}\Pi_i\left(\alpha_{k_i}^{(s_i)}-\beta_{k_i}^{(s_i)}\right)\notag\\
&\quad \times\biggl[\biggl(\alpha_{k_1}^{(s_1)*}\alpha_{k_2}^{(s_2)*}\alpha_{k_3}^{(s_3)*}+\beta_{k_1}^{(s_1)*}\beta_{k_2}^{(s_2)*}\beta_{k_3}^{(s_3)*}\biggr)\frac{1}{K^3}\biggl(1+3\frac{k_1k_2+k_2k_3+k_1k_3}{K^2}+15\frac{k_1k_2k_3}{K^3}\biggr)\notag\\
&\quad -\frac{1}{48}c_h^6k_1k_2(k_1+k_2)\eta_0^6\biggl(\alpha_{k_1}^{(s_1)*}\alpha_{k_2}^{(s_2)*}\beta_{k_3}^{(s_3)*}+\beta_{k_1}^{(s_1)*}\beta_{k_2}^{(s_2)*}\alpha_{k_3}^{(s_3)*}\biggr)\biggr]F(s_i,k_i)\biggr\}\biggr|_{\tilde k\to0}.
\end{align}
The integral which sets the $\eta_0$-dependence of the bispectrum is
\begin{align}
\int\D\eta\eta^5 e^{ic_h\tilde k\eta}.
\end{align}
Here, the Lagrangian in Eq.~(\ref{eq: ex-L}) is included in the general spatially covariant theory in Ref.~\cite{Gao:2014soa}. In this framework, the primordial tensor bispectrum in the presence of only the positive frequency mode has been calculated in Ref.~\cite{Akita:2015mho}. By choosing ${\rm Re}[\alpha^{(s)}_k]=1, {\rm Im}[\alpha^{(s)}_k]=0$, and $\beta^{(s)}_k=0$, one can see that our results reproduce those in Ref.~\cite{Akita:2015mho}. We also note that the resultant bispectra with the Bunch-Davies initial state vanish for the flattened triangles $\tilde k_j=0$ since $F(s_i,k_i)\propto\tilde k_j$ and suppressed around $\tilde k_j=0$.
\end{widetext}

\bibliography{Dth}

\end{document}